\documentclass[twocolumn,superscriptaddress,preprintnumbers,amsmath,amssymb,prl,floatfix]{revtex4-1}

\pdfoutput=1
\usepackage{graphicx} 
\usepackage{dcolumn} 
\usepackage{bm}
\usepackage{color}
\usepackage{hyperref}
\usepackage{graphicx}
\usepackage{amsmath}
\usepackage{setspace}
\usepackage{CJKutf8}

\begin{document}
\begin{CJK*}{UTF8}{gbsn}

\preprint{RIKEN-iTHEMS-Report-22}

\title{Calibration of nuclear charge density distribution by back-propagation neural networks}

\author{Zu-Xing Yang (杨祖星)}
\affiliation{RIKEN Nishina Center, Wako 351-0198, Japan}

\author{Xiao-Hua Fan (范小华)}
\affiliation{School of Physical Science and Technology, Southwest University, Chongqing 400715, China}

\author{Tomoya Naito (内藤智也)}
\affiliation{RIKEN Interdisciplinary Theoretical and Mathematical Sciences Program, Wako 351-0198, Japan}
\affiliation{Department of Physics, Graduate School of Science, The University of Tokyo, Tokyo 113-0033, Japan}

\author{Zhong-Ming Niu (牛中明)}
\affiliation{School of Physics and Optoelectronic Engineering, Anhui University, Hefei 230601, China}

\author{Zhi-Pan Li (李志攀)}
\affiliation{School of Physical Science and Technology, Southwest University, Chongqing 400715, China}

\author{Haozhao Liang (梁豪兆)}
\affiliation{Department of Physics, Graduate School of Science, The University of Tokyo, Tokyo 113-0033, Japan}
\affiliation{RIKEN Interdisciplinary Theoretical and Mathematical Sciences Program, Wako 351-0198, Japan}


\begin{abstract}
Based on the back-propagation neural networks and density functional theory, a supervised learning is performed firstly to generate the nuclear charge density distributions. 
The charge density is further calibrated to the experimental charge radii by a composite loss function.
It is found that, when the parity, pairing, and shell effects are taken into account, about $96\%$ of the nuclei in the validation set fall within two standard deviations of the predicted charge radii.
The calibrated charge density is then mapped to the matter density, and further mapped to the binding energies according to the Hohenberg-Kohn theorem. 
It provides an improved description of some nuclei in both binding energies and charge radii.
Moreover, the anomalous overbinding in $^{48}$Ca implies the existence of an indispensable beyond-mean-field effect.
\end{abstract}

\maketitle

\end{CJK*}


{\it Introduction}---The charge density distribution, experimentally given by electron scattering experiments \cite{Angeli2013At.DataNucl.DataTables99.6995, Helm1956Phys.Rev.104.14661475,Hofstadter1956Rev.Mod.Phys.28.214254}, is essential for extracting nuclear structure information, including the shell-structure evolution, shape coexistence, shape transition, and neutron-skin thickness \cite{Campbell2016Prog.Part.Nucl.Phys.86.127180, Wood1992Phys.Rep.215.101201, Bender2003Rev.Mod.Phys.75.121180, Li2016Nucl.Sci.Tech.27.}.
The relative radii of neighboring nuclei can also be determined using muonic-atom spectra as well as isotope shifts of laser spectroscopy \cite{Angeli2013At.DataNucl.DataTables99.6995}.
Since a strong connection exists between charge densities and matter densities, one generally calculates the charge density distribution by merging the nucleon form factors \cite{Dreher1974Nucl.Phys.A235.219248, Friar1975Adv.Nucl.Phys..219376} with the matter densities determined by scattering experiments, the ($-1p$) reaction \cite{Mutschler2016Nat.Phys.13.152156}, or even heavy-ion collisions \cite{Fan2019Phys.Rev.C99.041601}.

Early on, the Fermi distributions \cite{Yennie1954Phys.Rev.95.500512, Ullah1994Pramana43.165168} and Fourier-Bessel expansion \cite{Euteneuer1978Nucl.Phys.A298.452476,Ullah1994Pramana43.165168} were used to describe nuclear matter and charge densities approximately.
With the development of computational power, various branching models \cite{Bardeen1957Phys.Rev.106.162164, Bogoljubov1958Fortschr.Phys.6.605682, Dytrych2007Phys.Rev.Lett.98.162503, Dean2008Phys.Rev.Lett.101.119201, Roth2009Phys.Rev.C79.064324,  Dytrych2013Phys.Rev.Lett.111.252501, Neff2015Phys.Rev.C92.024003, Zhang2017Phys.Rev.C95.014316, Naito2019Phys.Rev.C99.024309, Bertulani2019Phys.Rev.C100.015802} derived from density functional theory (DFT) and shell model calculation have become more popular among theorists.
However, due to the complexity of nuclear many-body systems, the calculations of these theories still face challenges in describing the beyond-mean-field effects and nucleon-nucleon correlations \cite{Yang2019Phys.Rev.C100.054325, Tanaka2021Science371.260264, Duer2018Nature560.617621, Miller2019Phys.Rev.Lett.123.232003}.

For nuclear complex systems, back-propagation neural networks \cite{Rumelhart1986Nature323.533536} have achieved a series of success in various aspects, such as nuclear masses \cite{Niu2018Phys.Lett.B778.4853, Ma2020Phys.Rev.C101.045204, Athanassopoulos2004Nucl.Phys.A743.222235}, nuclear spins and parities \cite{Gernoth1993Phys.Lett.B300.17}, charge radii \cite{Utama2016J.Phys.GNucl.Part.Phys.43.114002, Wu2020Phys.Rev.C102.054323, Dong2022Phys.Rev.C105.014308, Cotextquotesingle2022Phys.Rev.C105.034320}, excited states \cite{Lasseri2020Phys.Rev.Lett.124.162502, Wang2022Phys.Lett.B830.137154}, extrapolation problems in an $ab~initio$ method \cite{Negoita2019Phys.Rev.C99.054308}, $\alpha$-decay half-lives \cite{Saxena2021J.Phys.GNucl.Part.Phys.48.055103}, $\beta$-decay half-lives \cite{Niu2019Phys.Rev.C99.064307}, fission yields \cite{Wang2019Phys.Rev.Lett.123.122501, Qiao2021Phys.Rev.C103.034621}, and so on. 
Most present neural networks learn and predict the residuals between theoretical and experimental values, which has an advantage that the corrected predictions are more accurate than the existing theoretical model \cite{Niu2018Phys.Lett.B778.4853, Utama2016Phys.Rev.C93.014311,Niu2019Phys.Rev.C99.064307}. 
However, the corrections have weak physical interpretability and it is difficult to predict different observables consistently.
Google proposed a hybrid quantum-classical machine learning model for training beyond classical data types, where back-propagation is used to tune the quantum logic gate parameters, allowing for deep integration of physics and neural networks \cite{Broughton2020.}.

Based on the Hohenberg-Kohn maps \cite{Moreno2020Phys.Rev.Lett.125.076402} of DFT and the proven strong generalization ability of neural networks in describing density distributions \cite{Yang2021Phys.Lett.B823.136650}, in this letter, we collectively constrain the charge density distributions by back-propagation with experimental data of charge radii, which makes the residual information flow back from radii to densities.
We also construct the map from charge density to binding energy to achieve further transfer of information from radius to binding energy.


{\it The charge density generators}---Previously, a multilayer feed-forward neural network with a back-propagation algorithm of error has been elaborated to perform the maximum likelihood estimation in the process of generating density distributions \cite{Yang2021Phys.Lett.B823.136650} approximating the theoretical calculation. 
It is shown that a network trained by the density distributions of about 200--300 nuclei is sufficient to describe the density distributions of all the nuclei on the nuclear chart and has a powerful extrapolation capability \cite{Yang2021Phys.Lett.B823.136650}.


\begin{figure}
\includegraphics[width=7 cm]{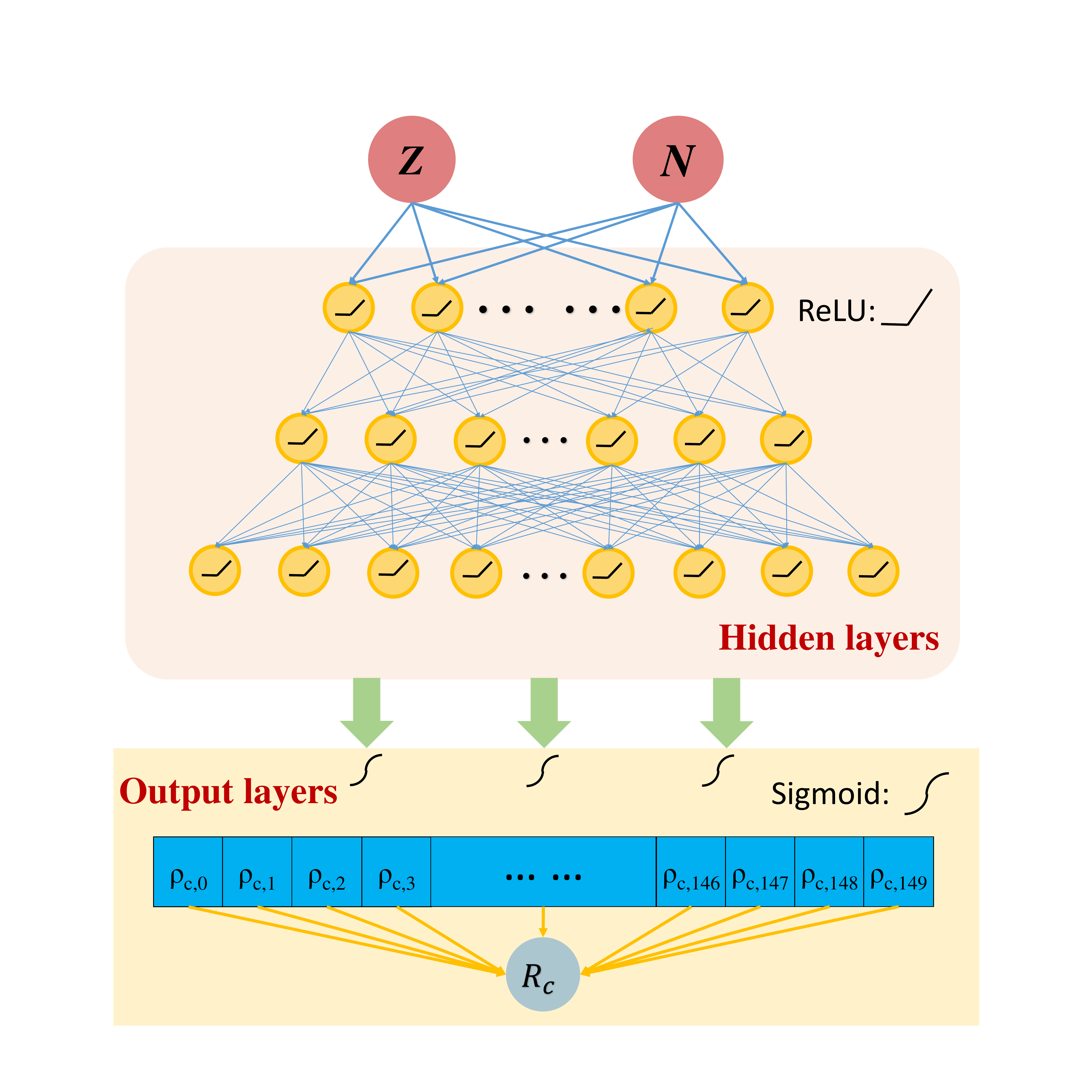}
\caption{\label{fig:net} (Color online) Schematic diagram of the structure of the charge density generator neural networks.}
\end{figure}

Based on such a high computational efficiency and generalization ability, in this study we introduce the correlation between the theoretical charge density and the experimental root-mean-square charge radius ($R_\text{c}$) to a new hybrid neural network.
The structure of this network is shown in Fig.~\ref{fig:net}, where $\rho_{\text{c},i} = \rho_\text{c}(r_i)$ is the density on the mesh with $r_i = 0.1 \times i ~\rm{fm}$ ($i=0,1,\ldots,149$). 
The input is the proton and neutron numbers of a nuclide, $\mathbf{x} = \{Z, N\}$.
The outputs are the charge density $\rho_\text{c}$ and root-mean-square charge radius $R_\text{c}$ gained by the customized integration layer $\hat{l}_{\rho_\text{c} \rightarrow R_\text{c}}$,
\begin{equation}
\hat{l}_{\rho_\text{c} \rightarrow R_\text{c}} : R_\text{c,pre} =\sqrt{\frac{\int \rho_\text{c,pre} r^4 dr}{\int \rho_\text{c,pre} r^2 dr}}, 
\end{equation}
where $\rho_\text{c,pre}$ is the normalized predicted charge density distribution, i.e., $\int 4\pi \rho_\text{c,pre}(r) r^2 d r = Z$.
For the hidden module, three attempts are made: 1. A deep fully connected neural network; 2. A convolutional neural network, suitable for learning the gradient information of features; 3. A neural network with a feature layer $\hat{l}_{i2\to i8}$, containing the parity, paring, and shell effects,
\begin{equation}
\hat{l}_{i2\to i8} :   \{Z,N\} \to  \{Z,N,\mathcal{O}_p,\mathcal{O}_n, \delta, v_p, v_n, \mathcal{P}\}  
\end{equation}
where
\begin{equation}
\begin{aligned}
&\mathcal{O}_p = Z\mod 2, &&\mathcal{O}_n = N\mod 2,\\
&\delta = [(-1)^Z+(-1)^N]/2, &&\mathcal{P}=v_pv_n/(v_p+v_n),
\end{aligned}
\end{equation}
and $v_p$ ($v_n$) is the difference between the actual nucleon number $Z$ ($N$) and the nearest magic number (8, 20, 28, 50, 82, or 126).
Hereafter, we refer to the three schemes of neural networks $\mathcal{M}$ as charge density generator-1 (CDG-1), CDG-2, and CDG-3, respectively.
See Supplemental Materials (SM) for details.
The map is denoted as
\begin{equation}
(\rho_\text{c} , R_\text{c})_\text{pre} =  \mathcal{M}(\mathbf{x},\mathbf{w}),
\end{equation}
where $\mathbf{w}$ is the set of trainable parameters.


In general, the charge density distribution calculated by physical models $\rho_\text{c,theo}$, e.g., that calculated by the Skyrme Hartree-Fock (SHF) theory with the Bardeen-Cooper-Schrieffer (BCS) pairing, is expected to be quite accurate.
The $R_\text{c}$ residuals between theory and experiment can be eliminated by assuming a correction to the charge density  $\delta\rho_\text{c}$, satisfying $\int_0^\infty \delta\rho_\text{c}(r) r^2\, dr = 0$.
We refer this process as calibration.
The calibrated charge density distribution $\rho_\text{c,cali}$, whose $R_\text{c}$ is expected to be close to $R_\text{c,exp}$ \cite{Angeli2013At.DataNucl.DataTables99.6995}, can be obtained by
\begin{equation}
\label{eq1}
\rho_\text{c,cali}(r) = \rho_\text{c,theo}(r)+ \delta\rho_\text{c}(r).
\end{equation}
Additionally, we aim at {\it making the smallest possible corrections $\delta\rho_\text{c}$ to the theory.}

To this end, we design a composite loss function.
The normalized mean-square-error $L_\rho$ \cite{Yang2021Phys.Lett.B823.136650} is employed as an assessment of density distribution: 
\begin{equation}
L_\rho = \frac{1}{N_\text{g}} \sum_{i=0}^{149} [\rho_\text{c,pre}(r_i)-\rho_\text{c,theo}(r_i)]^2 \times 1\,{\rm fm}^{6}, \end{equation}
where $N_\text{g} = 150$ indicates the number of grid points and the factor $1\,{\rm fm}^{6}$ makes $L_\rho$ dimensionless. 
The charge density distributions $\rho_\text{c,theo}(r)$ are calculated by
the SHF+BCS theory with the SkM* interaction \cite{Bartel1982Nucl.Phys.A386.79100}.
They are obtained from the charge form factor $F_{\text{c}}$ by the inverse Fourier-Bessel transform,
\begin{equation}
\label{eq:rhocth}
\rho_\text{c,theo}(r)=\frac{1}{2 \pi^{2}} \int dk \, k^{2} j_{0}(k r) F_{\text{c}}(k;\rho_n, \rho_p,...).
\end{equation}
where $j_0$ is the spherical Bessel function. 
The contributions of matter density and spin-orbit current are folded in $F_{\text{c}}$ \cite{Reinhard1991.2850} (see SM for details).
$R_\text{c}$ has a large range of variation (about $1$--$6\,\mathrm{fm}$) and therefore the Pearson $\chi^2$ divergence is picked, i.e., its loss function $L_\text{R}$ reads
\begin{equation}
L_\text{R} = \frac{( R_\text{c,pre}-R_\text{c,exp})^2}{R_\text{c,pre}} \times 1\,{\rm fm}^{-1},
\end{equation}
where the factor $1\,{\rm fm}^{-1}$ also makes $L_\text{R}$ dimensionless. 
The two loss functions $L_\rho$ and $L_\text{R}$ are combined as
\begin{equation}
\mathop{\mathrm{Loss}}(\mathcal{W},\mathbf{w})\equiv\frac{1}{B_s} \sum_{\text{n}_\text{u}=1}^{B_s} [(1-\mathcal{W}) L^{(\text{n}_\text{u})}_\rho(\mathbf{w}) + \mathcal{W} L^{(\text{n}_\text{u})}_\text{R}(\mathbf{w})],
\end{equation}
where $\mathcal{W}$ is the weighting factor and $B_s=64$ is batch size, which means that 64 nuclei are constrained simultaneously for each training session.
Actually, the corrections for different nuclei are derived from the same parameter updates $\delta \mathbf{w}$, i.e.,{\it ~the correction of each nucleus is uniformly constrained by the other nuclei}.
We randomly take 640 nuclei (10 batches) of about 900 nuclei measured to date as the training set. 
The remaining nuclei are recorded as the validation set (see SM for details).
It is recorded as an epoch when all nuclei on the training set have been trained once.


\begin{figure}
\includegraphics[width=8.5 cm]{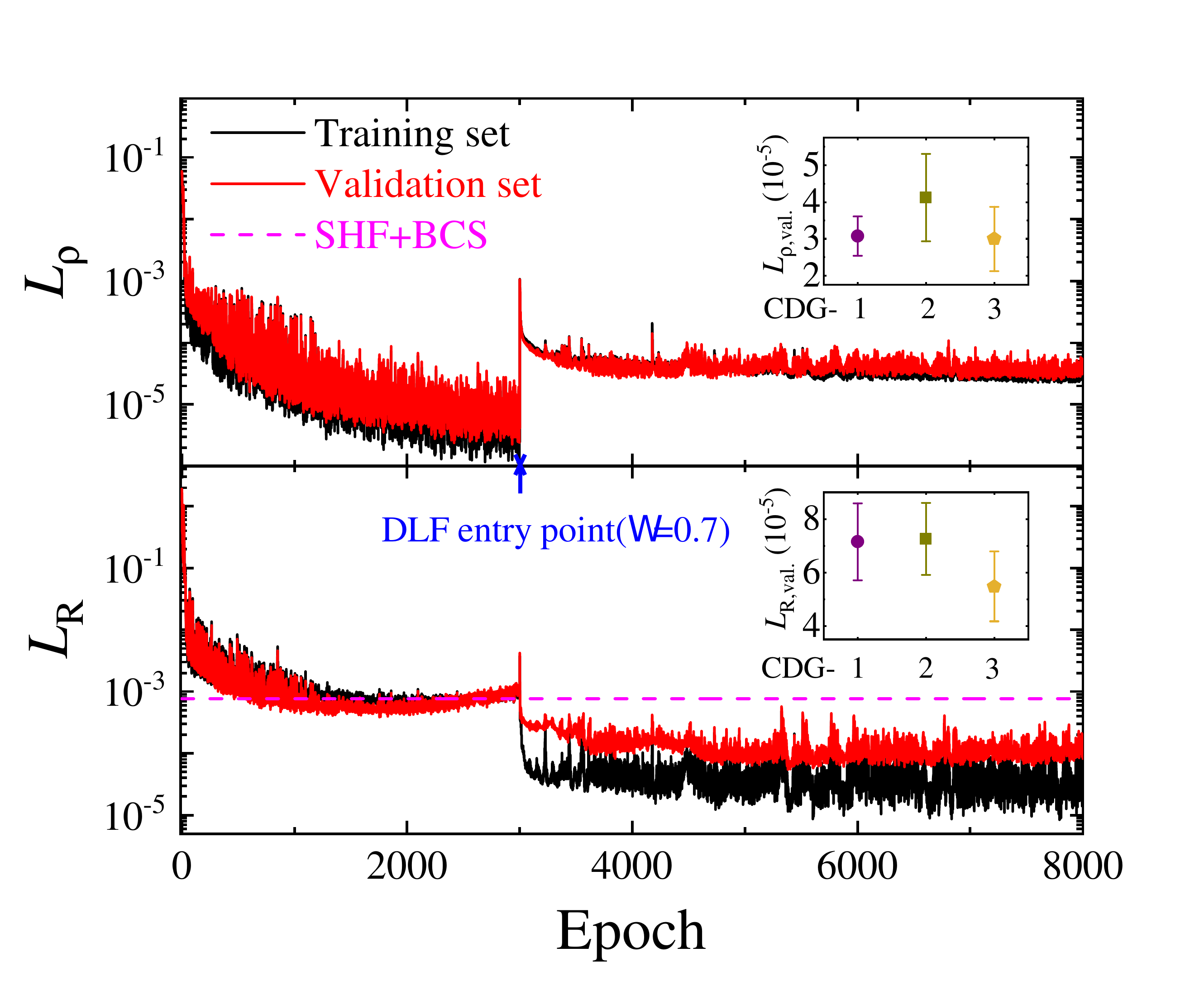}
\caption{\label{fig:loss} (Color online)
Upper panel: Normalized mean square error for density distributions on the training and validation sets as a function of epochs. Lower panel: Pearson $\chi^2$ divergence for charge radii as a function of epochs. 
Inserts: The loss values on validation set (upper) $L_{\rho,\text{val}}$ and (lower) $L_\text{R,val}$ for CDG-1, CDG-2, and CDG-3 with $\mathcal{W}=0.7$.}
\end{figure}

{\it Machine learning processes}---The process of machine learning is divided into two stages: simulating the SHF+BCS results and correcting with experimental data.
The evaluation of these processes is shown in Fig.~\ref{fig:loss}.
The first 3000 epochs are the stage of simulating the SHF+BCS results, during which $\mathop{\mathrm{Loss}}(0,\mathbf{w})=L_\rho$ is being minimized.
One can see that the loss functions of the training and validation sets almost overlap.
It means neither $\rho_c$ nor $R_c$ is overfitted, which shows the generalization ability of the network.
The dashed magenta line is the Pearson $\chi^2$ divergence $L_\text{R,theo}$ on the validation set between the SHF+BCS and experimental values:
\begin{equation}
L_\text{R,theo}=\frac{1}{N_\text{v}} \sum_{{\text{n}_\text{u}} \in val.} \frac{( R_\text{c,theo}^{(\text{n}_\text{u})}-R_\text{c,exp}^{(\text{n}_\text{u})})^2}{R_\text{c,theo}^{(\text{n}_\text{u})}}  \times 1\,{\rm fm}^{-1} ,
\end{equation}
where $N_\text{v}$ is the number of nuclei on the validation set.
After a short training, the loss values $L_\text{R}$ overlap with $L_\text{R,theo}$.
This indicates that the network naturally captures the $R_\text{c}$ information well in the process of learning the density distribution.

After 3000 epochs, the pre-trained model is further tuned with an objective function $\mathop{\mathrm{Loss}}(\mathcal{W}=0.7)$, which allows the importance of experimental data to slightly exceed that of theoretical calculations.
Thus, the charge density distributions under constraints are spontaneously corrected by the network.
Since the correction makes the predicted charge radius close to the experimental value and the distribution naturally deviate from the model, the value of $L_\text{R}$ falls and the value of $L_\rho$ jumps as shown in Fig.~\ref{fig:loss}.
It is clear that machine learning successfully eliminated a portion of residuals between SHF+BCS and experimental values.
Such a process can be migrated to any other theoretical model.
Remarkably, the training costs only 10 GPU minutes.

{\it Calibrated charge density and charge radius}---To explore the network performance, the errors of the densities and radii given by the CDG-1, CDG-2, and CDG-3 on the validation set are plotted in the inserts of Fig.~\ref{fig:loss}.
It can be seen that the errors by CDG-3 are minimal for both density and radius.
This indicates the predictions of CDG-3 are closest to the experimental radii, while its corrections to the theory are smallest.
This agrees with our assumption.
Compared to CDG-1 and CDG-2, CDG-3 takes into account the odd-even staggering, pairing, and shell effects.
Therefore, we conclude that the network structures considering more physical properties are more sufficient to improve the prediction accuracy, which is consistent with the mass research \cite{Niu2018Phys.Lett.B778.4853}.

\begin{figure}
\includegraphics[width=8.5 cm]{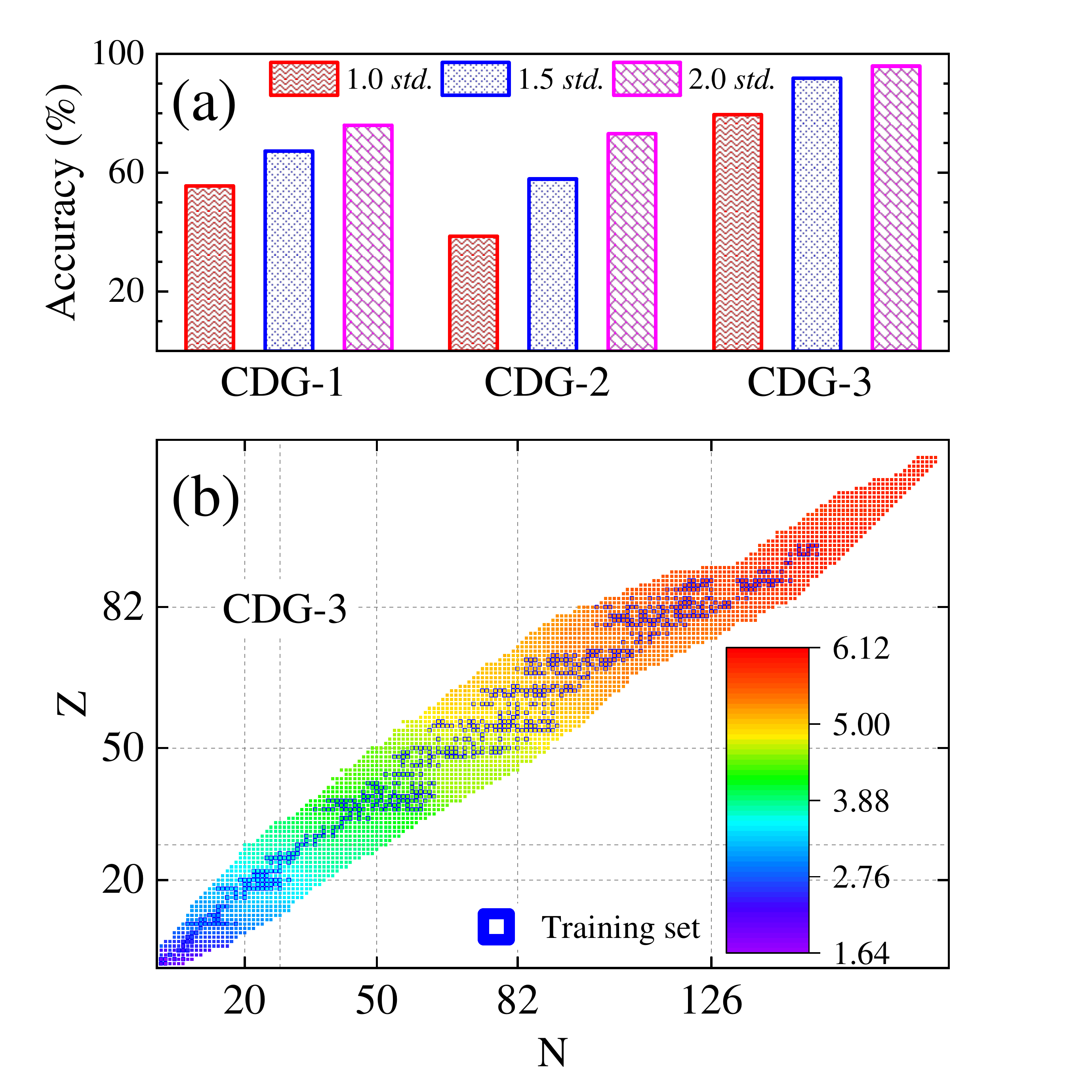}
\caption{\label{fig:map2} (Color online)
(a) The accuracy of neural networks CDG-1, CDG-2, and CDG-3, expressed by the percentages of the experimental data on the validation set falling within different predicted standard deviations (red for 1 $std.$, blue for 1.5 $std.$, and pink for 2 $std.$).
(b) Charge radii predicted by CDG-3, where the nuclei in the training set are denoted with the blue squares.}
\end{figure}

Accuracy is more intuitive to show the CDGs' performance than error. 
Figure~\ref{fig:map2}(a) shows the prediction accuracy of the three networks, where the accuracy indicates the percentages of the experimental data on the validation set falling within different predicted standard deviations (red for 1 $std.$, blue for 1.5 $std.$, and  pink for 2 $std.$).
The values of accuracy are consistent with the errors on validation set shown in Fig.~\ref{fig:loss} for different networks, i.e., the smaller the error, the higher the accuracy.
In particular, the numbers of experimental data falling within 2 standard deviations do not exceed $80\%$ for CDG-1 and CDG-2. 
Meanwhile, the accuracy of CDG-3 (1 $std.$) is already close to $80\%$, which even reaches $96\%$ for 2 $std.$.
It can be speculated that CDG-3 can predict the radii of the remaining $\sim 2000$ unmeasured nuclei with high precision.
We present the predictions for nuclear charge radii in Fig.~\ref{fig:map2}(b).

\begin{figure}
\includegraphics[width=8.5 cm]{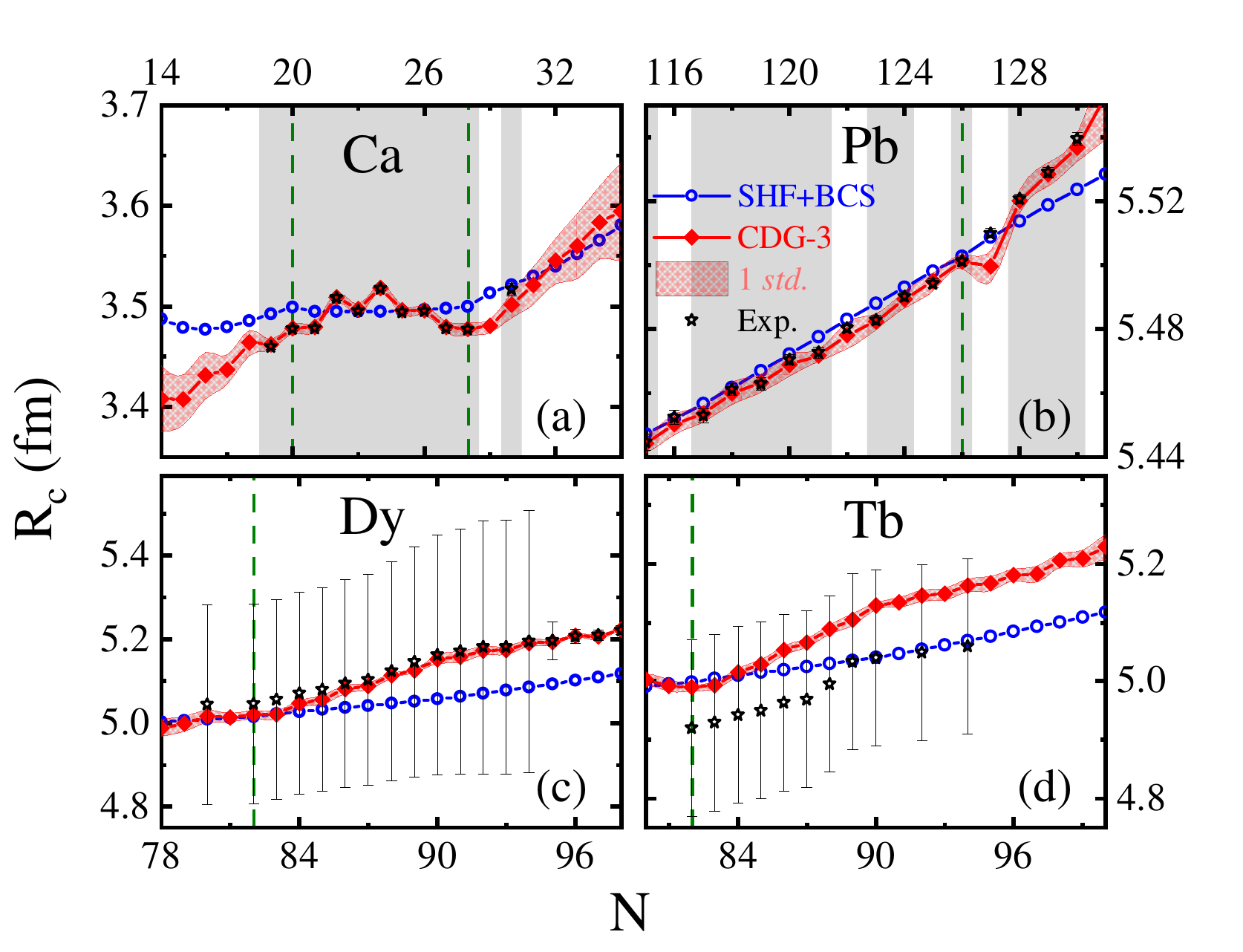}
\caption{\label{fig:radii2} (Color online)
Charge radii predicted by CDG-3 with 1-$std.$ error for the Ca, Pb, Dy, and Tb isotopes, where the training regions are indicated by shadows and the magic numbers are indicated by the vertical dashed lines.
The SHF+BCS results and experimental data are also shown for comparison.}
\end{figure}

By taking several isotopic chains as examples, we compare the charge radii calculated by SHF+BCS and the predictions of CDG-3 with 1-$std.$ error with the available experimental data in Fig.~\ref{fig:radii2}. 
It is found that CDG-3, containing the parity, paring, and shell effects, describes well the odd-even staggering for the Ca isotopes. 
Meanwhile, CDG-1 and CDG-2 failed in benchmarking the experimental data (see SM).
For Pb isotopes, CDG-3 performs comparably with the theoretical model.
The present predictions almost match the experimental data in the case of training several Pb isotopes.
In particular, the predictive power remains strong even when we remove the Ca and Pb isotopes from the training set (see SM).
Given the excellent performance of CDG-3, we predicted the untrained Dy and Tb isotopes.
For the Dy isotopes, the predictions are in agreement with experiment, but with much smaller uncertainty than the experimental error.
For the Tb isotopes, compared to the SHF+BCS results, the predicted isotope shifts are in good agreement with the experimental data, although the predicted charge radii are systematically larger than the present experimental central values by $\sim 0.1$~fm.
The above predictions can be testified in the coming experiments.

\begin{figure}
\includegraphics[width=6 cm]{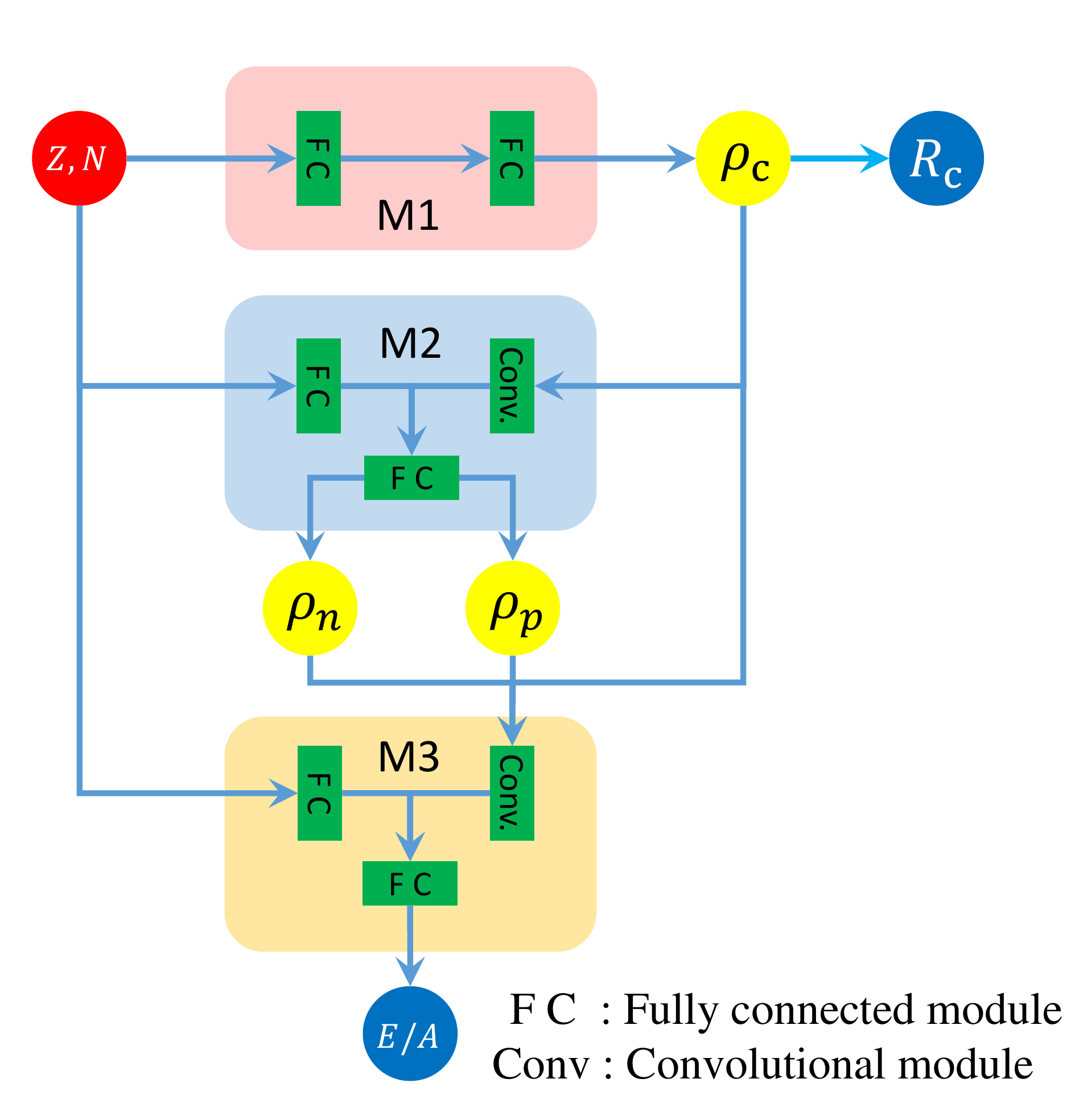}
\caption{\label{fig:net2} (Color online) Schematic diagram of the structure of feedforward neural network for the CDTBE map. }
\end{figure}

{\it Map from charge density to binding energy}---According to the Hohenberg-Kohn theorem \cite{Hohenberg1964Phys.Rev.136.B864B871, Moreno2020Phys.Rev.Lett.125.076402}, there exists a bijective map between the local matter density and the one-body potential.
This means that the corrections on the density distributions can be mapped to other observables, such as binding energies \cite{Ryczko2019Phys.Rev.A100.022512}.
Thus, we reconstruct a map from charge density to binding energy (CDTBE).
The schematic structure of such a feedforward neural network is shown in Fig.~\ref{fig:net2}. 
There are three parts in the CDTBE map---M1, M2, and M3.
M1 is the previously trained charge density generator, and we take the CDG-3;
M2 is a map from charge density to matter density as the inverse of Eq.~(\ref{eq:rhocth});
M3 is a map from the neutron, proton, and charge densities to the binding energy per nucleon.
M2 and M3 are trained by the densities and binding energies of SHF+BCS with effective interactions SkM, SkM*, SkIII, SLy4, SkT, and SkT3 \cite{Friedrich1986Phys.Rev.C33.335351} (See SM for details).
It is found that the networks adequately captures the relationship between densities and binding energy with different DFT effective interactions.
As the charge density updates in CDG-3, the residual information flows to other observables, i.e., $\delta R_\text{c} \to \delta \rho_\text{c} \to \{ \delta\rho_{n} , \delta\rho_{p} \} \to \delta(E/A)$.

\begin{figure}
\includegraphics[width=8 cm]{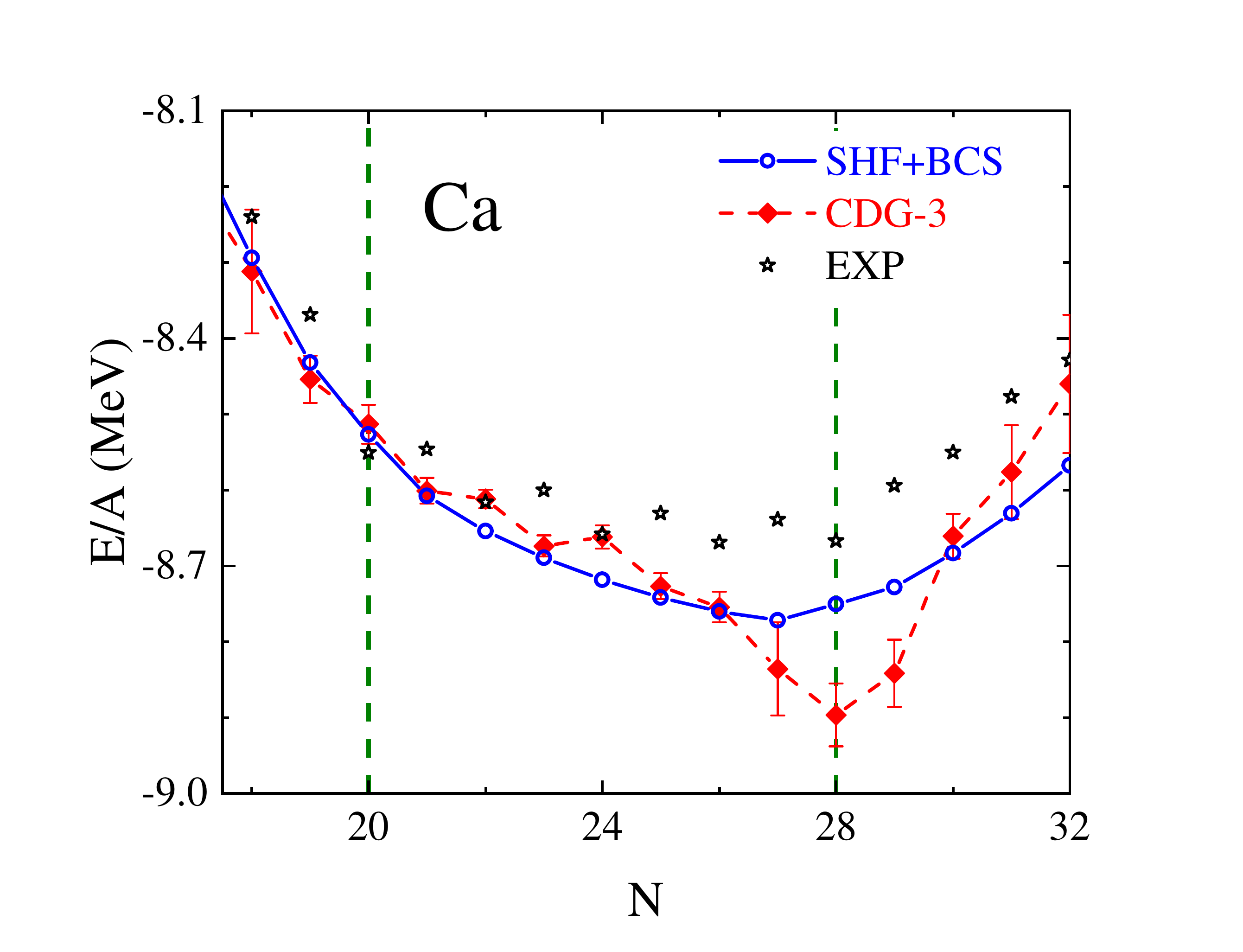}
\caption{\label{fig:skin} (Color online) Binding energy per nucleon of Ca isotops with statistical error (1 $std.$) obtained by CDG-3 residual information flow. The SHF+BCS results and experimental data are also shown.}
\end{figure}

Taking the Ca isotopes as examples, the binding energies per nucleon obtained by the CDG-3 residual information flow are shown in Fig.~\ref{fig:skin}.
We note that the propagated corrections provide a better description, except $^{47,48,49}$Ca.
Especially for $^{40,42,44}$Ca, both radii (see Fig.~\ref{fig:radii2}(a)) and binding energies coincide well with the experimental values.
This illustrates that the physical features contained in the CDTBE map based on density functional theory are adequate for these nuclei.
In contrast, for $^{48}$Ca, the radius is corrected to a smaller value than that of SHF+BCS, leading to an increase of nuclear densities in the central region and a further decrease of binding energy, which is consistent with equation of state.
However, it results in further deviations from the experimental value, which implies an indispensable beyond-mean-field effect near $^{48}$Ca, as discussed in Ref.~\cite{Perera2021Phys.Rev.C104.064313}.


{\it Summary}---In this study, a novel supervised learning on the combination of the theoretical charge density distributions and the experimental charge radii has been performed.
In such a way, the physics features embedded in nuclear density functional theory can be preserved to a large extent and the feedback from the experimental data can be considered quantitatively.
It is found that the description of charge radii can be improved globally on the nuclear chart.
In particular, the specific charge radii evolution in the Ca isotopes can be well reproduced by taking the parity, pairing, and shell effects into account.
This property remains valid even when all the Ca isotopes are excluded in the learning set.
The predictive power is also shown with the charge radii in the Dy and Tb isotopes, where the present experimental uncertainties are much larger than the prediction uncertainties.

Inspired by the Hohenberg-Kohn theorem, an information flow from charge density to binding energy has also been constructed and investigated.
It is found that for the Ca isotopes the improvement in the description of charge radii can also be propagated to the improvement in the description of binding energies, except $^{47,48,49}$Ca.
The corresponding analysis implies the
existence of an indispensable beyond-mean-field effect near $^{48}$Ca.

Along this direction, we will not only improve the description of different nuclear observables consistently but also strengthen the interpretability of the supervised learning.

\begin{acknowledgements}
{\it Acknowledgements}---This work is supported by the National Natural Science Foundation of China under Grants Nos.~12005175 and 11875070,
the Fundamental Research Funds for the Central Universities under Grant No.~SWU119076,
the JSPS Grant-in-Aid for Early-Career Scientists under Grant No.~18K13549,
the JSPS Grant-in-Aid for Scientific Research (S) under Grant No.~20H05648,
and the Anhui project (Z010118169).
This work is also partially supported by the RIKEN Pioneering Project: Evolution of Matter in the Universe, the RIKEN Special Postdoctoral Researchers Program, 
and the Science and Technology Hub Collaborative Research Program from RIKEN Cluster for Science, Technology and Inovation Hub (RCSTI).
\end{acknowledgements}

\bibliographystyle{apsrev4-1}
\bibliography{ref}

\end{document}


\beginsupplement
\preprint{RIKEN-iTHEMS-Report-22}

\begin{CJK*}{UTF8}{gbsn}

\title{Supplemental Material for ``Calibration of nuclear charge density distribution by back-propagation neural networks''}

\author{Zu-Xing Yang (杨祖星)}
\affiliation{RIKEN Nishina Center, Wako 351-0198, Japan}

\author{Xiao-Hua Fan (范小华)}
\affiliation{School of Physical Science and Technology, Southwest University, Chongqing 400715, China}

\author{Tomoya Naito (内藤智也)}
\affiliation{RIKEN Interdisciplinary Theoretical and Mathematical Sciences Program, Wako 351-0198, Japan}
\affiliation{Department of Physics, Graduate School of Science, The University of Tokyo, Tokyo 113-0033, Japan}

\author{Zhong-Ming Niu (牛中明)}
\affiliation{School of Physics and Optoelectronic Engineering, Anhui University, Hefei 230601, China}

\author{Zhi-Pan Li (李志攀)}
\affiliation{School of Physical Science and Technology, Southwest University, Chongqing 400715, China}

\author{Haozhao Liang (梁豪兆)}
\affiliation{Department of Physics, Graduate School of Science, The University of Tokyo, Tokyo 113-0033, Japan}
\affiliation{RIKEN Interdisciplinary Theoretical and Mathematical Sciences Program, Wako 351-0198, Japan}


\maketitle
\end{CJK*}

\section{Details of charge density generators}

In this section, in addition to Fig.~1, we explain the details of the three charge density generators (CDGs).
The corresponding networks are built through the following 6 kinds of layer operators.

\begin{enumerate}
\item {\bf Input layer} $\mathbf{x}$:
\begin{equation}
\mathbf{x} = \{Z,N\},
\end{equation}
where $Z$ and $N$ are the proton and neutron numbers, respectively.
\item {\bf Feature layer} ($\hat{l}_{i2\to i8}$):
\begin{equation}
\{Z,N\} \to  \{Z,N,\mathcal{O}_p,\mathcal{O}_n, \delta, v_p, v_n, \mathcal{P}\},  
\end{equation}
with
\begin{equation}
\begin{aligned}
&\mathcal{O}_p = Z\mod2, &&\mathcal{O}_n = N\mod2,\\
&\delta = [(-1)^Z+(-1)^N]/2, &&\mathcal{P}=v_pv_n/(v_p+v_n),
\end{aligned}
\end{equation}
where $v_p$ ($v_n$) is the difference between the actual nucleon number $Z$ ($N$) and the nearest magic number, which is among $8$, $20$, $28$, $50$, $82$, and $126$.
\item {\bf Fully connected layer} ($\hat{l}_{f,g(x)}$):
\begin{equation}
\begin{aligned}
y_j^{(L)} &= \sum^{\mathrm{D}^{(L-1)}}_i w_{ij}^{(L)} x_{i}^{(L)} + b_{j}^{(L)}, \\
x^{(L+1)}_i &= g(y^{(L)}_i),
\end{aligned}
\end{equation}
where $x_{i}^{(L)}$ and $y_j^{(L)}$ are the inputs and outputs of the layer $L$, $\mathrm{D}^{(L)}$ is the number of output, $w_{ij}^{(L)}$ and $b_{j}^{(L)}$ are the weights and biases, and $g(x)$ denotes the activation function, respectively. In particular, $\hat{l}_{f,r}$ and $\hat{l}_{f,s}$ denote $g(x) = \mathrm{ReLU}(x) =\max(0,x)$ and $g(x) = \mathrm{Sigmoid}(x) = 1/(1+e^{-x})$, respectively.
\item {\bf Convolutional layer} ($\hat{l}_{c,g(x)}$) \cite{Simonyan2014.}:
\begin{equation}
\begin{aligned}
y_{kn}^{(L)} &=\sum_j^{C_\text{in}^{(L)}} \sum^{K_s^{(L)}}_{i} w_{ki}^{(L)} x_{j,S_t n + i}^{(L)} + b_{kn}^{(L)}, \\
x^{(L+1)}_{kn} &= g(y^{(L)}_{kn}),
\end{aligned}
\end{equation}
where $k \in [0, C_\text{out}^{(L)})$ and  $n \in[0, \lceil \mathrm{D}^{(L-1)}/S_t^{(L)} \rceil)$ with $\lceil a \rceil $ denoting the round up of $a$,   $C_\text{in}^{(L)}$ and $C_\text{out}^{(L)}$ denote the numbers of channels of input and output, respectively.
Here, $C_\text{out}^{(L)}$ is also equal to the number of convolution kernels $w_{ki}^{(L)}$ in layer $L$.
Moreover, $S_t^{(L)}$ denotes the stride when the convolution kernel slides and $K_s^{(L)}$ denotes the convolution kernel size.
In this study, the one-dimensional convolution is used.
\item {\bf Pooling layer} ($\hat{l}_{p}$):
\begin{equation}
x^{(L+1)}_{n} = \max(x^{(L)}_{S_tn},x^{(L)}_{S_tn+1},...,x^{(L)}_{S_t n+K_s-1}),
\end{equation}
where $S_t^{(L)}$ and $K_s^{(L)}$ denote the stride and size, respectively. 
\item {\bf Radius layer} ($\hat{l}_{\rho \to R}$):
\begin{equation}
R_\text{pre} = \left[\frac{\int \rho_\text{pre}(r) r^4\, dr}{\int \rho_\text{pre}(r) r^2\, dr}\right]^{\frac{1}{2}}. 
\end{equation}
Through this layer, the density on the meshed $\rho_\text{pre}$ is transformed into the root-mean-square radius $R_\text{pre}$.
The back-propagation of this layer is achieved with the help of {\bf Autograd} in Pytorch \cite{Ketkar2017.195208}.
\end{enumerate}

\begin{table}
\centering
\caption{The structure and hyperparameters of CDG-1.}
\begin{tabular}{cccc} 
\hline
\hline
L & Types and Operators                   & D & $g(x)$   \\ 
\hline
0              & Input Layer ($\mathbf{x}$)                         & 2                 &               ---         \\
1               & Dense Layer ($\hat{l}_{f,r}$)                         & 50                & ReLU                   \\
2             & Dense Layer ($\hat{l}_{f,r}$)                         & 100               & ReLU                   \\
3              & Dense Layer ($\hat{l}_{f,r}$)                         & 200               & ReLU                   \\
4              & Dense Layer ($\hat{l}_{f,r}$)                         & 400               & ReLU                   \\
5              & Dense Layer ($\hat{l}_{f,r}$)                         & 800               & ReLU                   \\
6              & Dense Layer ($\hat{l}_{f,r}$)                         & 400               & ReLU                   \\
7              & Dense Layer ($\hat{l}_{f,r}$)                         & 200               & ReLU                   \\
8              & Output Layer ($\hat{l}_{f,s}$)                        & 150               & Sigmoid                \\ 
9             &  Radius Layer ($\hat{l}_{\rho \to R}$)                 & 1               & ---               \\ 
\hline
\multicolumn{4}{c}{Other information}         \\
\hline
\multicolumn{2}{c}{Numerical normalization factor}   & \multicolumn{2}{c}{1000/111}  \\
\multicolumn{2}{c}{Optimizer}   & \multicolumn{2}{c}{Adam}
\\
\multicolumn{2}{c}{Total parameters}   & \multicolumn{2}{c}{857,400}
\\
\multicolumn{2}{c}{Batch size}   & \multicolumn{2}{c}{64}
\\
\multicolumn{2}{c}{Running time}& \multicolumn{2}{c}{447.543 s}
\\
\hline
\hline
\end{tabular}
\label{ts1}
\end{table}

\begin{table}
\centering
\caption{The structure and hyperparameters of CDG-2.}
\begin{tabular}{cccccccc} 
\hline
\hline
 L & ~~~~~~~Types and Operators~~~~~~~                   & D &$C_\text{in}$ & $C_\text{out}$ & $K_s$ & $S_t$ & $g(x)$   \\ 
\hline
0              & Input Layer ($\mathbf{x}$)                         & 2  &--- & --- & --- & ---                &               ---         \\
1               & Dense Layer ($\hat{l}_{f,r}$)                         & 50    &--- &--- &--- & ---             & ReLU                   \\
2             & Dense Layer ($\hat{l}_{f,r}$)                         & 100 &--- & --- &--- & ---              & ReLU                   \\
3              & Dense Layer ($\hat{l}_{f,r}$)                         & 200  &--- & --- & --- & ---              & ReLU                   \\
4              & Dense Layer ($\hat{l}_{f,r}$)                         & 400  &--- & --- & --- & ---              & ReLU                   \\
5              & Dense Layer ($\hat{l}_{f,r}$)                         & 800  &--- & --- & --- &---              & ReLU                   \\
6              & Convolutional Layer ($\hat{l}_{c,r}$)                 & ---  &1 & 5 & 4 & 2              & ReLU                   \\
7              & Pooling Layer ($\hat{l}_{p}$)                         & ---  &--- & --- & 3 & 3              & ---                   \\
8              & Convolutional Layer ($\hat{l}_{c,r}$)                 & ---  &5 & 20 & 5 & 2              & ReLU                   \\
9              & Pooling Layer ($\hat{l}_{p}$)                         & ---  &--- & --- & 5 & 3              & ---                   \\
10              & Convolutional Layer ($\hat{l}_{c,r}$)                 & ---  &20 & 100 & 5 & 1              & ReLU                   \\
11              & Pooling Layer ($\hat{l}_{p}$)                         & ---  &--- & --- & 5 & 3              & ---                   \\
12              & Convolutional Layer ($\hat{l}_{c,r}$)                 & ---  &100 & 200 & 5 & 1              & ReLU                   \\
13              & Output Layer ($\hat{l}_{f,s}$)                        & 150 &--- & --- & --- & ---               & Sigmoid                \\ 
14              & Radius Layer ($\hat{l}_{\rho \to R}$)                        & 1 &--- & --- & --- & ---               & ---                \\ 
\hline
\multicolumn{8}{c}{Other information}  \\
\hline
\multicolumn{3}{c}{Numerical normalization factor}   & \multicolumn{5}{c}{1000/111}  \\
\multicolumn{3}{c}{Optimizer}   & \multicolumn{5}{c}{Adam}
\\
\multicolumn{3}{c}{Total parameters}   & \multicolumn{5}{c}{567,645}
\\
\multicolumn{3}{c}{Batch size}   & \multicolumn{5}{c}{64}
\\
\multicolumn{3}{c}{Running time}& \multicolumn{5}{c}{817.412 s}
\\
\hline
\hline
\end{tabular}
\label{ts2}
\end{table}

\begin{table}
\centering
\caption{The structure and hyperparameters of CDG-3.}
\begin{tabular}{cccc} 
\hline
\hline
 L & Types and Operators                   & D & $g(x)$   \\ 
\hline
0              & Input Layer ($\mathbf{x}$)                         & 2                 &               ---         \\
1              & Feature Layer ($\hat{l}_{i2\to i8}$)                         & 8                 &               ---         \\
2               & Dense Layer ($\hat{l}_{f,r}$)                         & 50                & ReLU                   \\
3             & Dense Layer ($\hat{l}_{f,r}$)                         & 100               & ReLU                   \\
4              & Dense Layer ($\hat{l}_{f,r}$)                         & 200               & ReLU                   \\
5              & Dense Layer ($\hat{l}_{f,r}$)                         & 400               & ReLU                   \\
6              & Dense Layer ($\hat{l}_{f,r}$)                         & 800               & ReLU                   \\
7              & Dense Layer ($\hat{l}_{f,r}$)                         & 400               & ReLU                   \\
8              & Dense Layer ($\hat{l}_{f,r}$)                         & 200               & ReLU                   \\
9              & Output Layer ($\hat{l}_{f,s}$)                        & 150               & Sigmoid                \\ 
10             &  Radius Layer ($\hat{l}_{\rho \to R}$)                 & 1               & ---               \\ 
\hline
\multicolumn{4}{c}{Other information}   \\
\hline
\multicolumn{2}{c}{Numerical normalization factor}   & \multicolumn{2}{c}{1000/111}  \\
\multicolumn{2}{c}{Optimizer}   & \multicolumn{2}{c}{Adam}
\\
\multicolumn{2}{c}{Total parameters}   & \multicolumn{2}{c}{857,700}
\\
\multicolumn{2}{c}{Batch size}   & \multicolumn{2}{c}{64}
\\
\multicolumn{2}{c}{Running time}& \multicolumn{2}{c}{450.277 s}
\\
\hline
\hline
\end{tabular}
\label{ts3}
\end{table}

The CDG-1, CDG-2, and CDG-3 are built by stacking the above unit layers and the corresponding details are listed in Tables~\ref{ts1}, ~\ref{ts2}, and ~\ref{ts3}, respectively.
These networks can also be denoted symbolically as
\begin{equation}
\begin{aligned}
\mathcal{M}_\text{CDG-1} &= \{\hat{l}_{f,r}^7, \hat{l}_{f,s}, \hat{l}_{\rho \to R} \}, \\
\mathcal{M}_\text{CDG-2} &= \{\hat{l}_{f,r}^5, (\hat{l}_{c,r}, \hat{l}_{p})^3 ,\hat{l}_{c,r}  \hat{l}_{f,s}, \hat{l}_{\rho \to R} \}, \\
\mathcal{M}_\text{CDG-3} &= \{\hat{l}_{i2 \to i8}, \hat{l}_{f,r}^7, \hat{l}_{f,s}, \hat{l}_{\rho \to R} \}.
\end{aligned}    
\end{equation}
We extend the 2-dimensional input to 800 dimensions by linear or non-linear operations, which will spontaneously generate more features in order to extract the information of density distribution. 
Subsequently, this high-dimensional information is descended to produce the density distribution $\rho_{\text{c},i}$.
During the calculations, the densities are multiplied by a numerical normalization factor, with which the densities are restricted between $0$ and $1$ with no effect on the root-mean-square radius. It is found that this can accelerate the speed of convergence \cite{Yang2021Phys.Lett.B823.136650}.
The Adaptive Momentum Estimation (Adam) \cite{Kingma2015.} is chosen as the optimizer for the stochastic gradient descent.
The network prediction slightly depends on the initialization of the network parameters $\mathbf{w}_\text{initial}$, which makes it possible to derive the statistical errors by initializing them differently.
All the trainable parameters are then determined, through the supervised learning of theoretical densities and the calibration based on experimental charge radii as stated in the main text.
In practice, all calculations are performed on the NVIDIA GeForce RTX 3070.

\section{Training and performance of CDGs}

In this section, we describe the training procedures and discuss the corresponding performance of CDG-1, CDG-2, and CDG-3 in detail.

\begin{figure}
\includegraphics[width=13.3 cm]{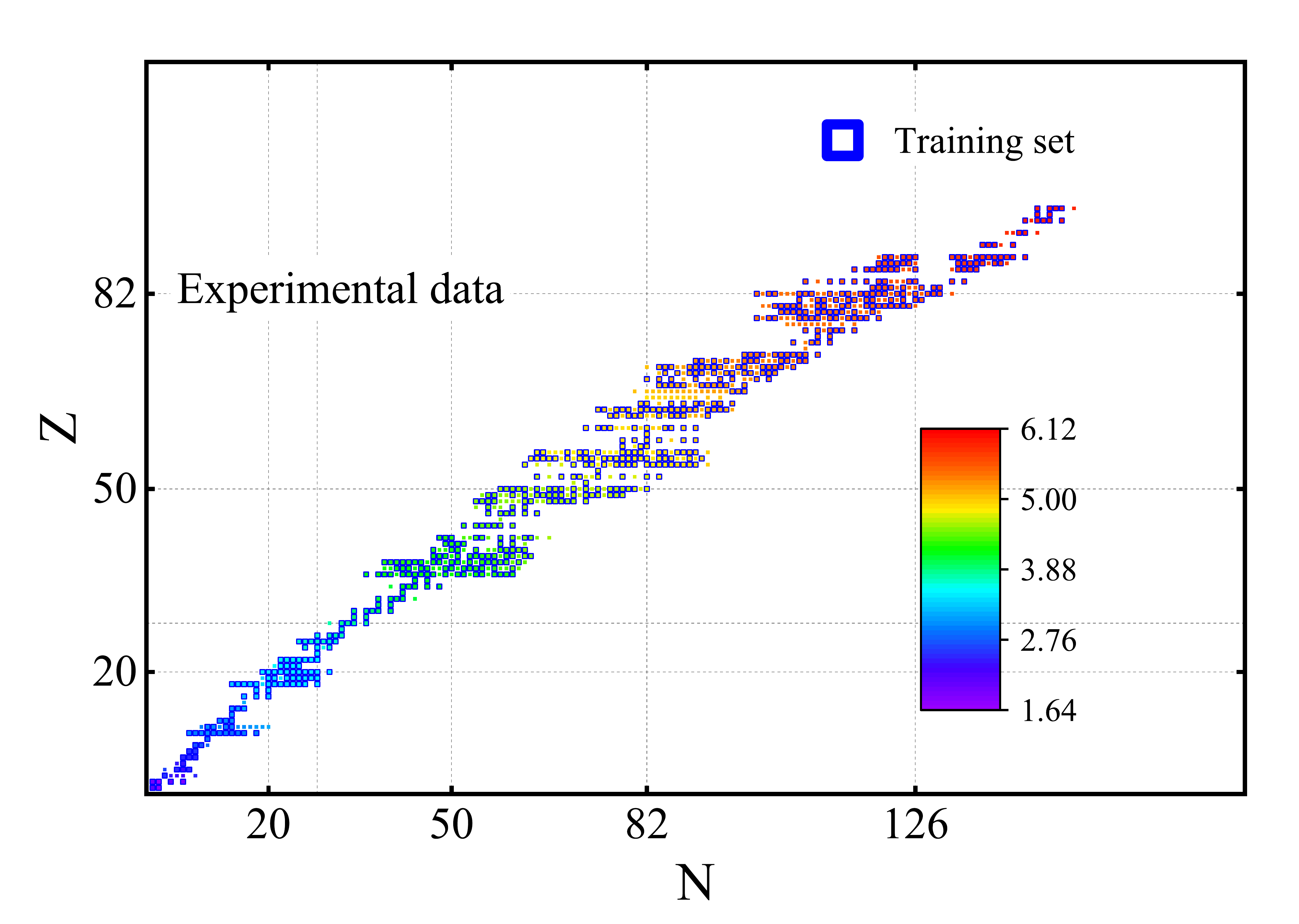}
\caption{\label{fig:exp} (Color online) Experimental data of charge radius, where the nuclei in the training set are denoted with the blue squares.}
\end{figure}

\begin{figure}
\includegraphics[width=13.3 cm]{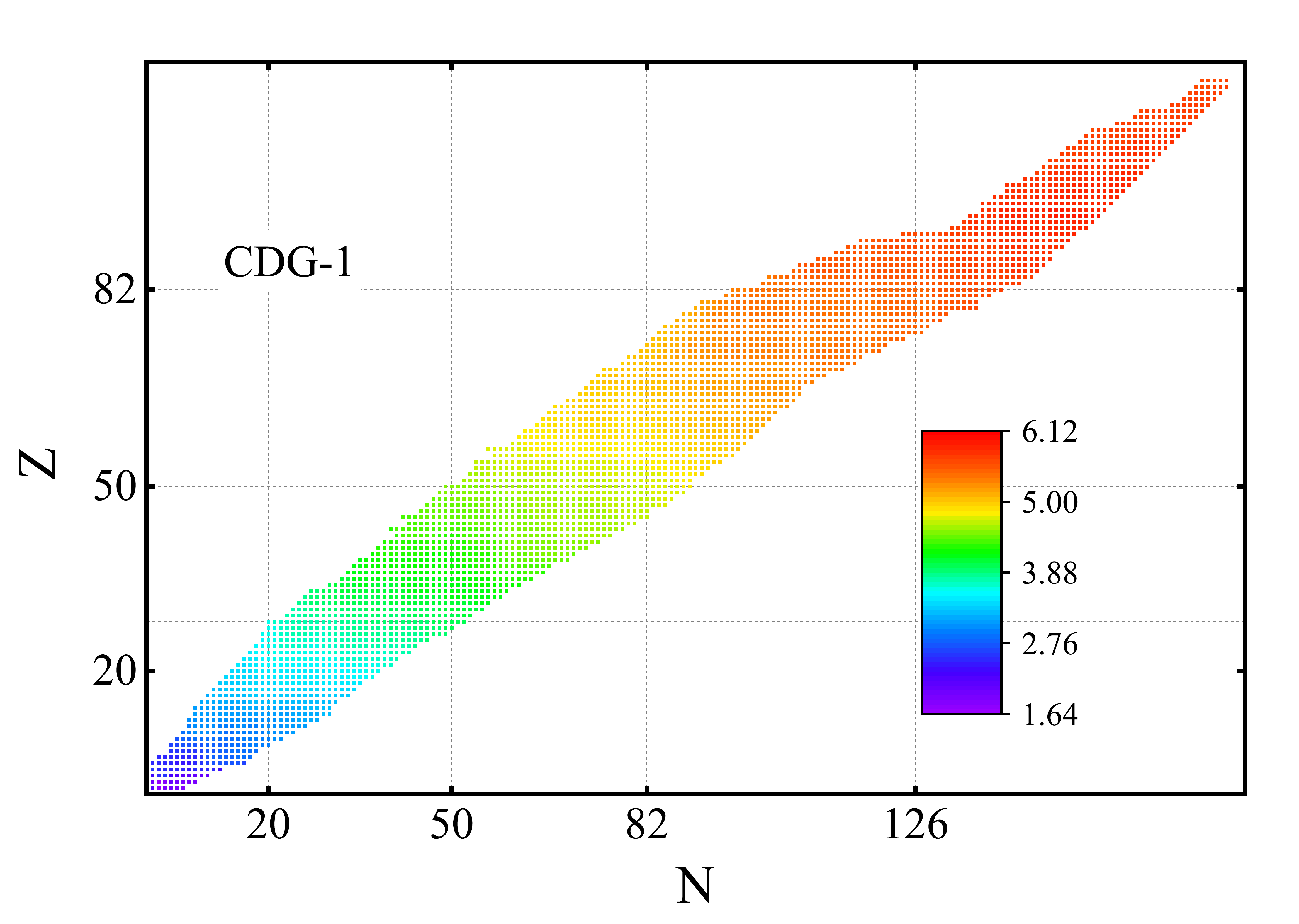}
\caption{\label{fig:cdg1} (Color online) Charge radii predicted by CDG-1.}
\end{figure}

\begin{figure}
\includegraphics[width=13.3 cm]{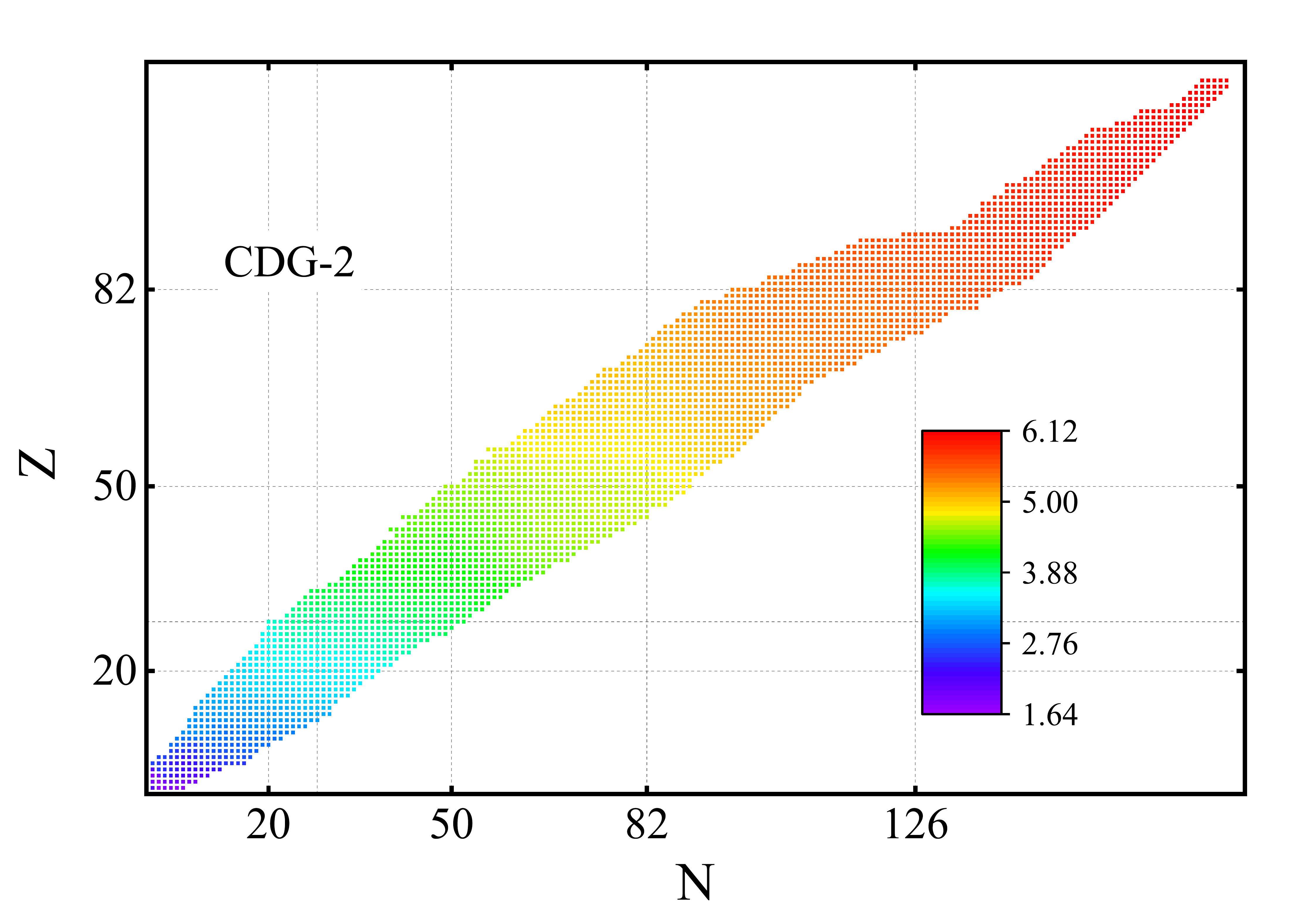}
\caption{\label{fig:cdg2} (Color online) Charge radii predicted by CDG-2.}
\end{figure}

\begin{figure}
\includegraphics[width=13.3 cm]{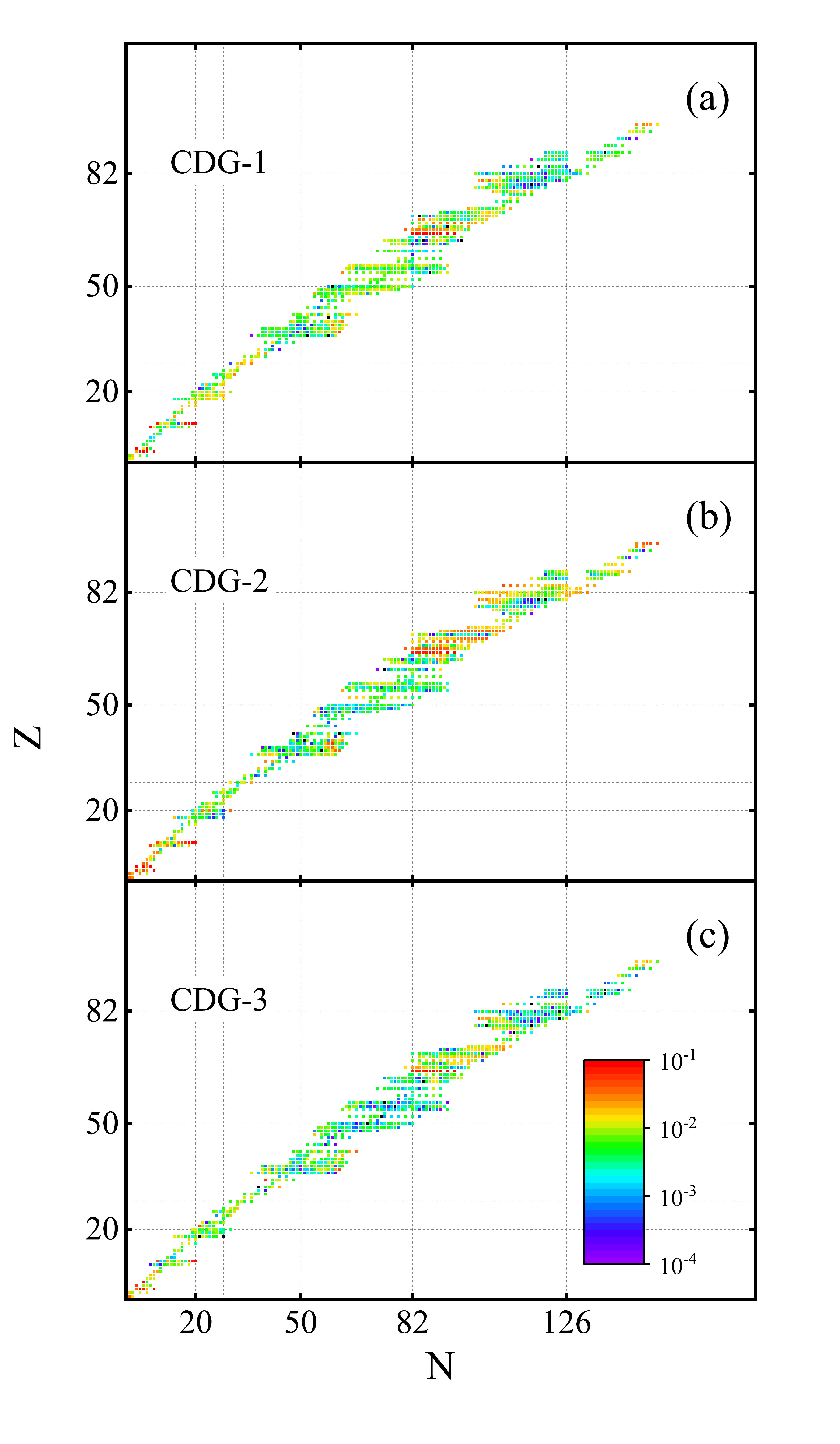}
\caption{\label{fig:cdg1diff} (Color online) The difference between the charge radii predicted by CDGs and experiential value.}
\end{figure}

The experimental data and the selected training nuclei are shown in Fig.~\ref{fig:exp}.
In addition to Fig.~3, the corresponding predicted charge radii by CDG-1 and CDG-2 are shown in Figs.~\ref{fig:cdg1} and \ref{fig:cdg2}, respectively.
The corresponding difference $\Delta R_\text{c}$ between the charge radii predicted by CDG-1, CDG-2, and CDG-3 and experiential value are shown in Fig.~\ref{fig:cdg1diff}.

\begin{figure}
\includegraphics[width=15 cm]{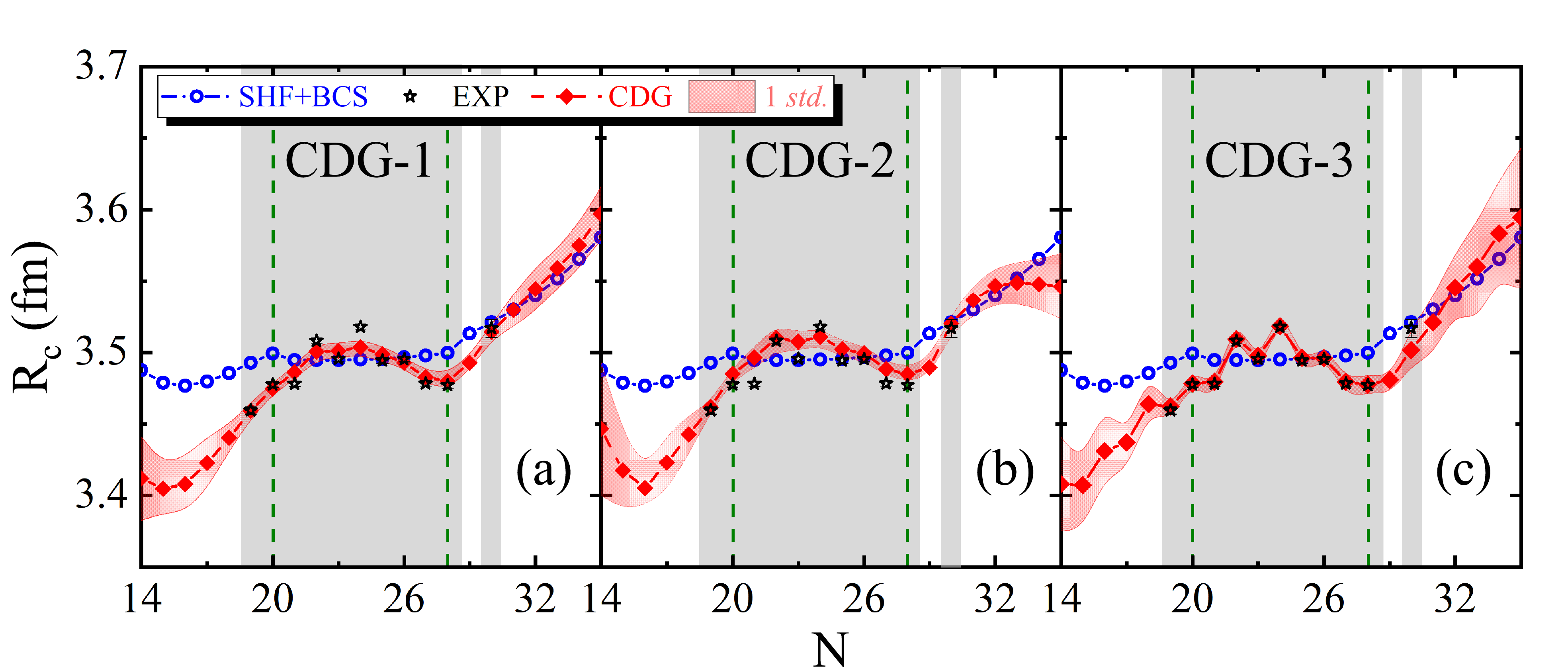}
\caption{\label{fig:CDG123} (Color online) The charge radii for Ca isotopes predicted by CDG-1, CDG-2, and CDG-3 with 1-$std.$ error, where the training regions are denoted by shadows and the magic numbers are denoted by the vertical dashed lines. The SHF+BCS results and experimental data are also shown for comparison.}
\end{figure}

In addition to Fig.~4, the charge radii for the Ca isotopes predicted by CDG-1, CDG-2, and CDG-3 are shown in Fig.~\ref{fig:CDG123}.
Noting that the training sets are identical for all CDGs, it can be seen that, although the overall tendency that $R_\text{c}$ increases from $^{39}$Ca to $^{44}$Ca, decreases from $^{44}$Ca to $^{48}$Ca, and then increases again from $^{48}$Ca to $^{50}$Ca can be in general described by all models, the delicate odd-even staggering in $R_\text{c}$ can only be described by CDG-3, which takes the parity, paring, and shell effects into account by the Feature Layer.

\begin{figure}
\includegraphics[width=10 cm]{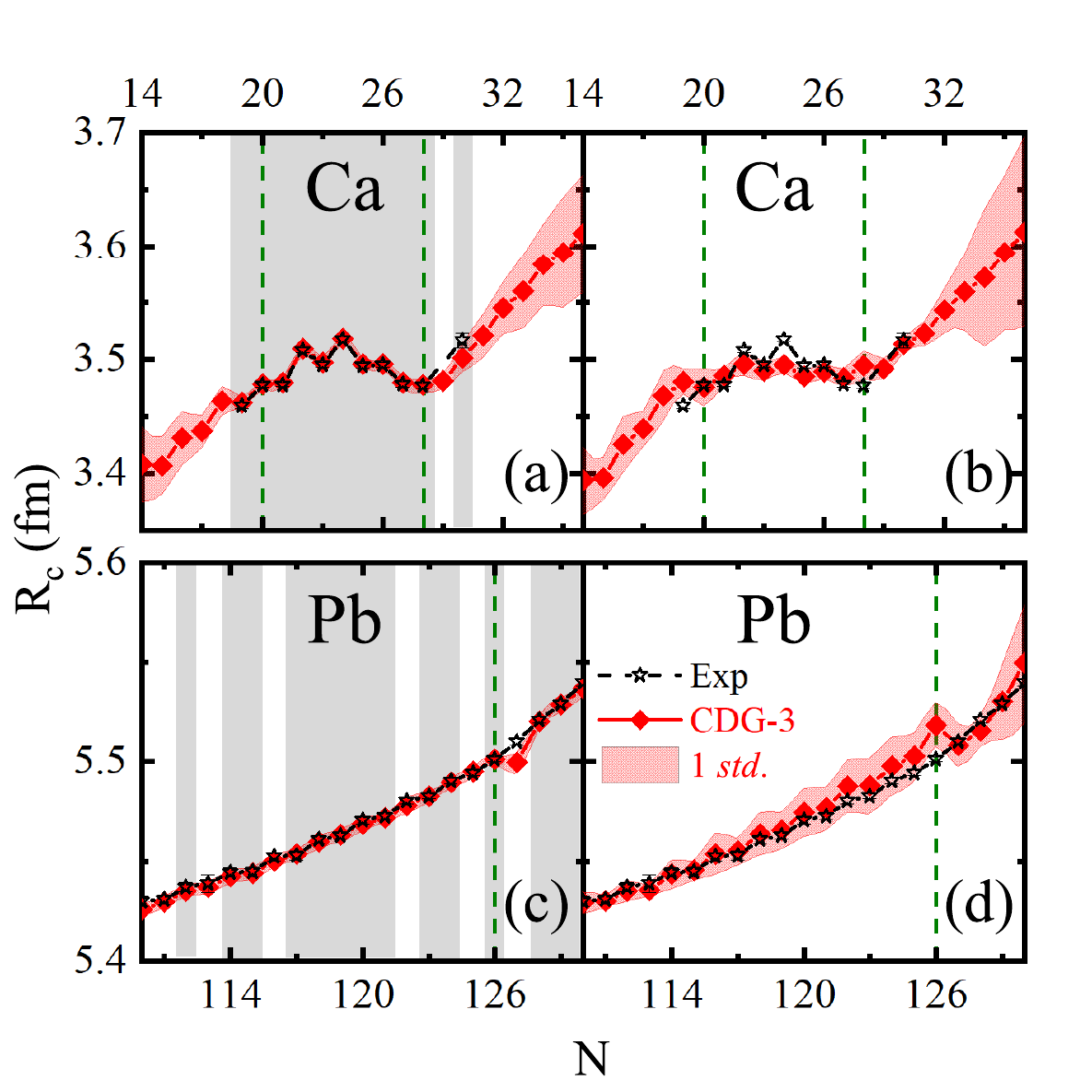}
\caption{\label{fig:radii2} (Color online) Predictions of charge radii for Ca by CDG-3 with the Ca isotopes (a) trained and (b) untrained, and predictions of charge radii for Pb by CDG-3 with the Pb isotopes (c) trained and (d) untrained, where the training regions are denoted by shadows and the magic numbers are denoted by the vertical dashed lines.}
\end{figure}

In order to further test the network predictive power, here we train CDG-3 with the training set that all the Ca isotopes are replaced by other random nuclei.
The corresponding predictions are shown in Fig.~\ref{fig:radii2}(b) compared with Fig.~\ref{fig:radii2}(a) that employs the same training set as shown in Fig.~\ref{fig:exp}.
It is seen that, although the whole Ca isotopic chain is untrained, only a few nuclei ($^{39}$Ca, $^{44}$Ca, and $^{48}$Ca) cannot be covered by one standard deviation of the prediction.
With the same idea, Figs.~\ref{fig:radii2}(c) and \ref{fig:radii2}(d) show the CDG-3 predictions for the Pb isotopes with and without training the Pb isotopes.
It is also clear that without training the predictions for the whole Pb isotopic train (except $^{208}$Pb) reproduce the experimental data within one standard deviation.

\begin{figure}
\includegraphics[width=18 cm]{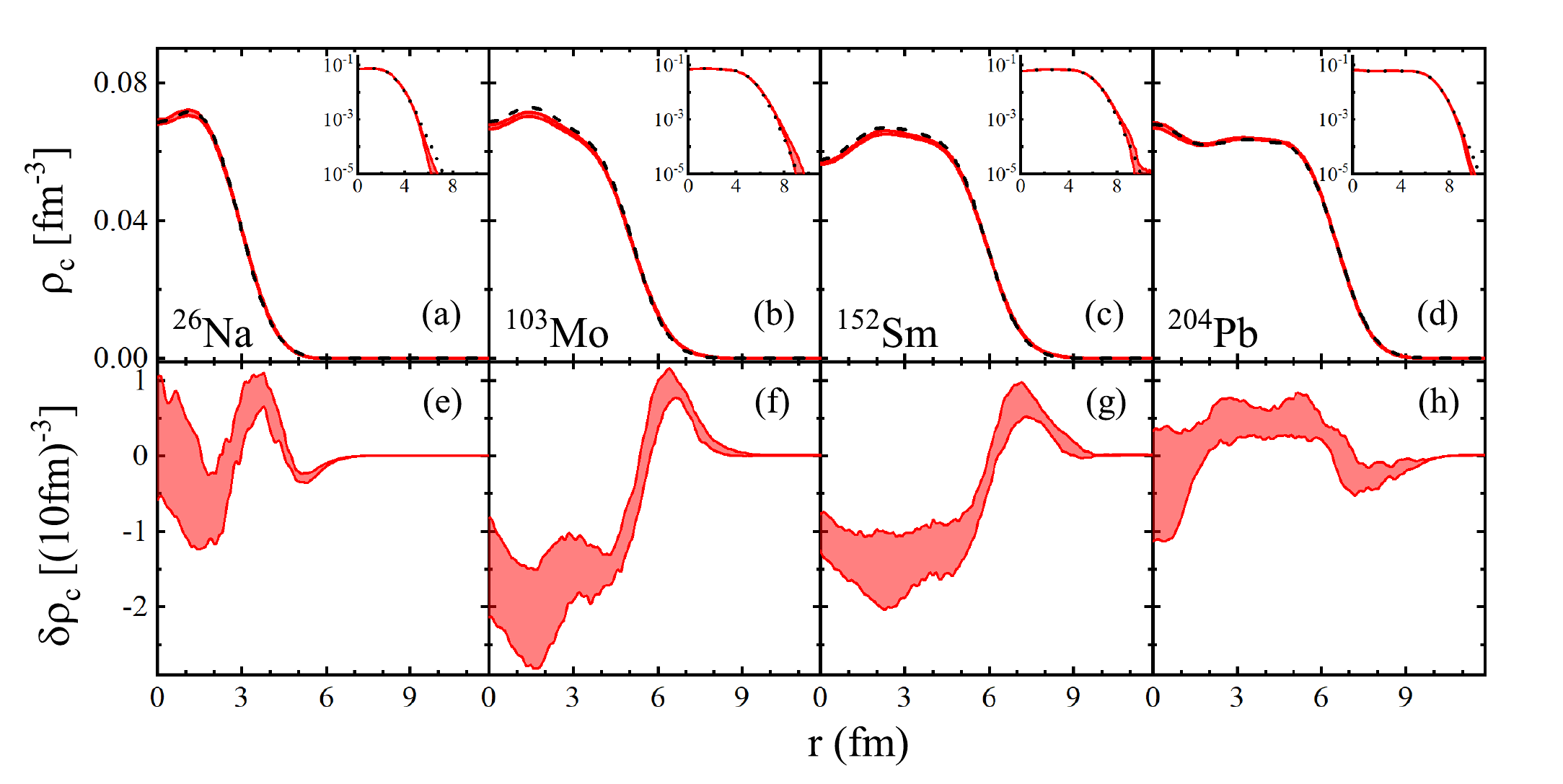}
\caption{\label{fig:profile} (Color online)
Upper panels: The charge density distributions of four untrained random nuclei---(a) $^{26}$Na, (b) $^{103}$Mo, (c) $^{152}$Sm, and (d) $^{204}$Pb---predicted by CDG-3, and compared with the original SHF+BCS results, where the inserts are in the logarithmic scales. Lower panels: The corresponding corrections $\delta \rho_c$ by CDG-3.}
\end{figure}

The radius residual information is transferred to the correction of the density distribution by
\begin{equation}
\label{eq:drho}
\begin{aligned}
\delta \rho, \delta R &=  \mathcal{M}(\mathbf{x},\mathbf{w}^{\mathrm{MLE}}_\text{theo} + \delta \mathbf{w}) - (\rho , R)_\text{theo}, \\
\delta \mathrm{w} &= \sum_{t_2} \mathcal{F} (\frac{\partial \mathop{\mathrm{Loss}}(0.7, \mathbf{w}^{(t_2)})}{\partial \mathrm{w}}),
\end{aligned}
\end{equation}
where $\mathcal{F}$ is the Adam optimizer.
The network weights $\mathbf{w}^{\mathrm{MLE}}_\text{theo}$ to describe the theoretical distribution are biased by $\delta \mathbf{w}$ due to the inclusion of experimental data.
The corresponding results are presented in Fig.~\ref{fig:profile}.
The charge density distributions of four untrained random nuclei---$^{26}$Na, $^{103}$Mo, $^{152}$Sm, and $^{204}$Pb---predicted by CDG-3 are shown and compared with the original SHF+BCS results, where the inserts are shown in the logarithmic scales.
As can be seen from Fig.~\ref{fig:profile}, the corrections are particularly small.
These fine corrections are of the magnitude of $\rm 10^{-3}$ $\rm fm^{-3}$ and converge within a very narrow interval.
For different nuclei, the corrections $\delta \rho_c$ convergence to different patterns, and the correction patterns are determined by $\delta \mathbf{w}$.
The diversity of such correction patterns may deserve further studies.


\section{Details of calculating charge densities in density functional theory}

In the SHF+BCS theory, we calculate the matter densities using the equation  $\rho_\tau(\mathbf{r}) = \Sigma_{i,\sigma} w_i \vert\phi_i(\mathbf{r},\sigma,\tau) \vert^2$, where $w_i$ denote the occupancy probability for each single-particle state and $\tau = p~(n)$ corresponds to proton (neutron) \cite{Reinhard1991.2850}. 
With the spherical symmetry, the single-particle wave function reads $\phi_i(\mathbf{r}) = R_i(r)/r \times \mathcal{Y}_{jlm}(\theta,\psi)$, where $\mathcal{Y}_{jlm}(\theta,\psi)$ is a spinor spherical harmonics.
Taking into account that the nucleon itself has an intrinsic electromagnetic structure, we need to fold the proton and neutron densities with the intrinsic charge density of nucleons.
Such a folding becomes a product in Fourier space, so we transform the densities into the so-called form factors $F_{\tau}(k)$,
\begin{equation}
F_{\tau}(k)=4 \pi \int_{0}^{\infty} \mathrm{d} r \, r^{2} j_{0}(k r) \rho_{\tau}(r),
\label{eqs1}
\end{equation}
where $j_0$ is the spherical Bessel function of the zeroth order. 
Similarly, the form factor of the spin-orbit current $\nabla \cdot \boldsymbol{J}$, accounting for the magnetic contributions to the charge density, is written as $F_{ls,\tau}(k)$.
The charge form factor is then given by
\begin{equation}\label{eqs2}
F_{\mathrm{c}}(k)= \sum_{\tau}\left[F_{\tau}(k) G_{\mathrm{E}, \tau}(k)+F_{l s, \tau}(k) G_{\mathrm{M}, \tau}(k)\right]
 \exp \left[\frac{(\hbar k)^{2}}{8\left\langle P_{\mathrm{cm}}^{2}\right\rangle}\right],
\end{equation}
where $G_{\mathrm{E}, \tau}(k)$ and $G_{\mathrm{M}, \tau}(k)$ are the electric and magnetic form factors of nucleons, respectively, and the exponential factor is the center-of-mass correction.
See Ref.~\cite{Reinhard1991.2850} for details.
Finally, the charge density is obtained from the charge form factor by the inverse Fourier-Bessel transform,
\begin{equation}\label{eqs3}
\rho_\text{c,theo}(r)=\frac{1}{2 \pi^{2}} \int_{0}^{\infty} \mathrm{d} k\, k^{2} j_{0}(k r) F_{\mathrm{c}}(k).
\end{equation}

\section{Details of CDTBE map}

In this section, in addition to Fig.~5, we explain the details of the feedforward neural network for the map from charge density to binding energy (CDTBE).

There are three parts in the CDTBE map---M1, M2, and M3.
M1 is the previously trained charge density generator, and we take the CDG-3; M2 is a map from charge density to matter density; M3 is a map from the neutron, proton, and charge densities to the binding energy per nucleon.


\begin{table}
\centering
\caption{The structure and hyperparameters of M2.}
\begin{tabular}{cccccccc} 
\hline
\hline
\multicolumn{8}{c} {Cell-1 ($\hat{C}_1 = \{ \hat{l}_{f,r}^3\}$)} \\
\hline
 L & Types and Operators                   & D & & & & & $g(x)$   \\ 
\hline
0              & Input Layer ($\mathbf{x}$)                            & 2  & & & &                &      ---         \\
1               & Dense Layer ($\hat{l}_{f,r}$)                        & 32      & & & &          & ReLU                   \\
2             & Dense Layer ($\hat{l}_{f,r}$)                          & 64      & & & &         & ReLU                   \\
3              & Dense Layer ($\hat{l}_{f,r}$)                         & 128       & & & &       & ReLU                   \\
\hline
\multicolumn{8}{c} {Cell-2 ($\hat{C}_2 = \{ \hat{l}_{c,r},\hat{l}_{p},\hat{l}_{c,r}, \hat{l}_{f,r}\} $) }\\
\hline
 L & ~~~~~~~Types and Operators~~~~~~~                   & D &$C_\text{in}$ & $C_\text{out}$ & $K_s$ & $S_t$ & $g(x)$   \\ 
\hline
0              & Input Layer ($\rho_c$)                    & 150   &--- & --- & --- & ---          &      ---         \\
1       & Convolutional Layer ($\hat{l}_{c,r}$)           & ---  &1 & 8 & 3 & 3              & ReLU                   \\
2   & Pooling Layer ($\hat{l}_{p}$)                         & ---  &--- & --- & 2 & 2              & ---                   \\
3       & Convolutional Layer ($\hat{l}_{c,r}$)           & ---  &8 & 32 & 5 & 5              & ReLU                   \\
4              & Dense Layer ($\hat{l}_{f,r}$)        & 512   &--- & --- & --- & ---              & ReLU                   \\
\hline
\multicolumn{8}{c} {Cell-3 ($\hat{C}_3= \{  \hat{l}_{f,r}^2 \} $) }\\
\hline
 L & ~~~~~~~Types and Operators~~~~~~~                   & D & &  &  &  & $g(x)$   \\ 
\hline
0               & Dense Layer ($\hat{l}_{f,r}$)                        & 768      & & & &          & ReLU                   \\
1               & Dense Layer ($\hat{l}_{f,r}$)                        & 1024      & & & &          & ReLU                   \\
\hline
\multicolumn{8}{c} {Cell-4 ($\hat{C}_4= \{  \hat{l}_{f,r}^2, \hat{l}_{f,s}\}$ ) }\\
\hline
 L & ~~~~~~~Types and Operators~~~~~~~                   & D & &  & & & $g(x)$   \\ 
\hline
0               & Dense Layer ($\hat{l}_{f,r}$)                        & 512      & & & &          & ReLU                   \\
1               & Dense Layer ($\hat{l}_{f,r}$)                        & 256      & & & &          & ReLU                   \\
2               & Dense Layer ($\hat{l}_{f,s}$)                        & 150      & & & &          & Sigmoid                  \\
\hline
\multicolumn{8}{c} {Cell-5 ($\hat{C}_5= \{  \hat{l}_{f,r}^2, \hat{l}_{f,s}\}$) }\\
\hline
 L & ~~~~~~~Types and Operators~~~~~~~                   & D & &  & & & $g(x)$   \\ 
\hline
0               & Dense Layer ($\hat{l}_{f,r}$)                        & 512      & & & &          & ReLU                   \\
1               & Dense Layer ($\hat{l}_{f,r}$)                        & 256      & & & &          & ReLU                   \\
2               & Dense Layer ($\hat{l}_{f,s}$)                        & 150      & & & &          & Sigmoid                  \\
\hline
\multicolumn{8}{c}{Other information}     \\
\hline
\multicolumn{3}{c}{Numerical normalization factor}   & \multicolumn{5}{c}{6.25}  \\
\multicolumn{3}{c}{Optimizer}   & \multicolumn{5}{c}{Adam}
\\
\multicolumn{3}{c}{Total parameters}   & \multicolumn{5}{c}{2,763,404}
\\
\multicolumn{3}{c}{Batch size}   & \multicolumn{5}{c}{32}
\\
\multicolumn{3}{c}{Running time}& \multicolumn{5}{c}{2241.901 s}
\\
\hline
\hline
\end{tabular}
\label{m2}
\end{table}

On the one hand, M2 is composed of 5 cells ($\hat{C}_{1,2,3,4,5}$), which can be regarded as the inverse of Eqs.~(\ref{eqs1}), (\ref{eqs2}), and (\ref{eqs3}).
We denote the model as
\begin{equation}
\left [
\begin{array}{c}
\hat{C}_4 \\
\hat{C}_5
\end{array}
\right ]
\left [
\hat{C}_3 (\hat{C}_1(\mathbf{x}) \uplus \hat{C}_2(\rho_c))
\right ] =
\left [
\begin{array}{c}
\rho_p \\
\rho_n
\end{array}
\right ].
\end{equation}
Here, the symbol $\uplus$ indicates splicing two vectors, e.g., $[1,3,2] \uplus [2,5,7] = [1,3,2,2,5,7]$.
The detailed structure of M2 is listed in Table~\ref{m2}.


\begin{table}
\centering
\caption{The structure and hyperparameters of M3.}
\begin{tabular}{cccccccc} 
\hline
\hline
\multicolumn{8}{c} {Cell-1 ($\hat{C}_1 = \{ \hat{l}_{f,r}^3\}$)} \\
\hline
 L & Types and Operators                   & D & & & & & $g(x)$   \\ 
\hline
0              & Input Layer ($\mathbf{x}$)                            & 2  & & & &                &      ---         \\
1               & Dense Layer ($\hat{l}_{f,r}$)                        & 32      & & & &          & ReLU                   \\
2             & Dense Layer ($\hat{l}_{f,r}$)                          & 64      & & & &         & ReLU                   \\
3              & Dense Layer ($\hat{l}_{f,r}$)                         & 128       & & & &       & ReLU                   \\
\hline
\multicolumn{8}{c} {Cell-2 ($\hat{C}_2 = \{ \hat{l}_{c,r},\hat{l}_{p},\hat{l}_{c,r}, \hat{l}_{f,r}\} $) }\\
\hline
 L & ~~~~~~~Types and Operators~~~~~~~                   & D &$C_\text{in}$ & $C_\text{out}$ & $K_s$ & $S_t$ & $g(x)$   \\ 
\hline
0              & Input Layer ($\rho_{p,n,c}$)                    & 150   &3 & --- & --- & ---          &      ---         \\
1       & Convolutional Layer ($\hat{l}_{c,r}$)           & ---  &3 & 24 & 3 & 3              & ReLU                   \\
2   & Pooling Layer ($\hat{l}_{p}$)                         & ---  &--- & --- & 2 & 2              & ---                   \\
3       & Convolutional Layer ($\hat{l}_{c,r}$)           & ---  &24 & 72 & 5 & 5              & ReLU                   \\
4              & Dense Layer ($\hat{l}_{f,r}$)        & 512   &--- & --- & --- & ---              & ReLU                   \\
\hline
\multicolumn{8}{c} {Cell-3 ($\hat{C}_3= \{  \hat{l}_{f,r}^2 \} $) }\\
\hline
 L & ~~~~~~~Types and Operators~~~~~~~                   & D & &  &  &  & $g(x)$   \\ 
\hline
0               & Dense Layer ($\hat{l}_{f,r}$)                        & 768      & & & &          & ReLU                   \\
1               & Dense Layer ($\hat{l}_{f,r}$)                        & 1024      & & & &          & ReLU                   \\
\hline
\multicolumn{8}{c} {Cell-4 ($\hat{C}_4= \{  \hat{l}_{f,r}^5, \hat{l}_{f,s}\}$ ) }\\
\hline
 L & ~~~~~~~Types and Operators~~~~~~~                   & D & &  & & & $g(x)$   \\ 
\hline
0               & Dense Layer ($\hat{l}_{f,r}$)                        & 1024      & & & &          & ReLU                   \\
1               & Dense Layer ($\hat{l}_{f,r}$)                        & 512      & & & &          & ReLU                   \\
2               & Dense Layer ($\hat{l}_{f,r}$)                        & 256      & & & &          & ReLU                   \\
3               & Dense Layer ($\hat{l}_{f,r}$)                        & 128      & & & &          & ReLU                 \\
4               & Dense Layer ($\hat{l}_{f,r}$)                        & 32      & & & &          & ReLU                   \\
5               & Dense Layer ($\hat{l}_{f,s}$)                        & 1      & & & &          & Sigmoid                  \\
\hline
\multicolumn{8}{c}{Other information}    \\
\hline
\multicolumn{3}{c}{Numerical normalization factor}   & \multicolumn{5}{c}{6.25}  \\
\multicolumn{3}{c}{Optimizer}   & \multicolumn{5}{c}{Adam}
\\
\multicolumn{3}{c}{Total parameters}   & \multicolumn{5}{c}{2,177,241}
\\
\multicolumn{3}{c}{Batch size}   & \multicolumn{5}{c}{32}
\\
\multicolumn{3}{c}{Running time}& \multicolumn{5}{c}{2740.545 s}
\\
\hline
\hline
\end{tabular}
\label{m3}
\end{table}

On the other hand, M3 is composed of 4 cells ($\hat{C}_{1,2,3,4}$).
We denote the model as
\begin{equation}
\hat{C}_4
\left [
\hat{C}_3 (\hat{C}_1(\mathbf{x}) \uplus \hat{C}_2(\rho_{p,n,c}))
\right ] = E/A.
\end{equation}
The detailed structure of M3 is listed in Table~\ref{m3}.

In the present study, M2 and M3 are trained by the densities and binding energies calculated by SHF+BCS with different effective interactions, SkM, SkM*, SkIII, SLy4, SkT, and SkT3.
As a result, the obtained learning deviations are less than $0.2\%$.

\clearpage

\bibliographystyle{apsrev4-1}
\bibliography{ref}